\documentstyle[emlines]{article}
\title{Quantum Relational Databases}
\author{Paul Cockshott}
\begin{document}
\maketitle
\def\dyad#1#2#3{(#1$\otimes$#2 $\rightarrow$ #3)}
\def\monad#1#2{(#1$\rightarrow$#2)}
\def\triad#1#2#3#4{(#1$\otimes$#2$\otimes$#3 $\rightarrow#4)$}

\def\dif{\mskip-\medmuskip \mkern5mu \mathbin{\bf dif}\penalty900 \mkern5mu \mskip-\medmuskip}
\def\proj{\mskip-\medmuskip \mkern5mu \mathbin{\bf proj}\penalty900 \mkern5mu \mskip-\medmuskip}
\def\join{\mskip-\medmuskip \mkern5mu \mathbin{\Join}\penalty900 \mkern5mu \mskip-\medmuskip}
\def\tuple#1{\langle {#1} \rangle}
\def\card#1{\# #1}
\def\map{{\cal M}}
\def\reduce#1#2{#1 \cal R #2}

\def\sel#1#2{#1{ \cal S\ #2 }}
Following the early suggestion by Deutsch \cite{Deutsch1985},
there has been considerable discussion in the
literature of the possibility of building quantum computing
machines. This has moved from basic discussion about the
concept of such machines through studies of 
the mathematical properties  of logic gates
that might be adequate to build them\cite{Deutsch1989} \cite{Feynman1986}
\cite{Deutsch1995}\cite{Sleator1995}, to discussions of practical algorithms
that might be run on them \cite{Shor1994}.
Despite doubts that have been expressed about the
physical practicability of quantum computers due to the problem of
decoherence \cite{Landauer1994}\cite{Shor1995}, there seems
good reason to hope that these are soluble\cite{DiVincenzo1995} in light of
the development of quantum error correcting codes\cite{Cipra1996}
\cite{Shor1997}.
Increasing numbers of practical suggestions for the technological
implementation of quantum computers have been advanced ranging
from the use of cold trapped ions \cite{Cirac1995} to the
use of NMR technology \cite{Gershenfeld1997}.

 Although conventional computers using semi-conductors
rely upon quantum effects in their underlying technology, their design
principles are classical. They have a definite state vector 
and they evolve deterministically between states in this space. Thus
the state of a classical computer with an $n$ bit store is defined
by a position in an $n$ dimensional binary co-ordinate state space. 

In contrast the state of
a quantum computer with a store made up of $n$ quantum two state
observables, or qubits, is given by a point in $2^n$ dimensional
Hilbert space. Each dimension of this space corresponds to one
of the $2^n$ possible values that $n$ classical bits can assume. 
These possible bit patterns constitute basis vectors for the Hilbert
space, and, associated with each such basis vector there is a complex
valued amplitude. At any instant the quantum computer is in a linear 
superposition
of all of its possible bit patterns.
It is this ability to exist in multiple states at once that is
exploited by algorithms such as Shor's method of prime factorisation
\cite{Shor1994}.

If we abstract from the difficult technical problem of 
long term coherent storage of qubit vectors, this ability of the store
to exist in multiple simultaneous states may be relevant to database
compression.

In the well established relational model\cite{Codd1970} data is
stored in  {\em relations} or tables.
Given sets $S_1, S_2, иии, S_n$, (not necessarily distinct), 
$R$ is a relation on these $n$ sets if it is a set of $n$-tuples each of which
has its first element from $S_1$, its second element from $S_2$, etc. 
The set $S_i$ is known as the $i$th {\em domain} of $R$.
Each row in a database table represents a tuple of the relation.
The tuples are conventionally represented as a vector of bits
divided into fields $F_1,F_2,...,F_n$ where $F_i$ contains
a symbol, drawn from some binary encoded alphabet, corresponding
to an element of $S_i$. If a single row can be encoded in 
$c$ bits and we have $r$ rows, then the whole databse
occupies $c.r$ bits.

In a quantum system a row could, using the same encoding, be
represented in $c$ qubits. However, use of superposition
of states would allow a single vector of $c$ qubits to 
represent all $r$ rows, each with an appropriate amplitude.
It is evident that were we to make a classical measurement
on such a superposited tuple, we would only be able to
read out one of the $r$ rows of the database. The measurement
would cause the wave function of the database to collapse
onto one of its tuples.

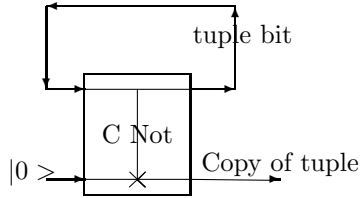
\begin{figure}\center{
\unitlength=1.00mm
\special{em:linewidth 0.4pt}
\linethickness{0.4pt}
\begin{picture}(43.03,30.03)
\put(17.00,5.00){\framebox(14.00,16.00)[cc]{C Not}}
\emline{17.00}{19.00}{1}{31.00}{19.00}{2}
\emline{17.00}{7.00}{3}{31.00}{7.00}{4}
\emline{24.00}{19.00}{5}{24.00}{7.00}{6}
\emline{25.00}{6.00}{7}{24.00}{7.00}{8}
\emline{23.00}{8.00}{9}{25.07}{6.02}{10}
\emline{25.07}{6.02}{11}{24.00}{6.91}{12}
\emline{24.00}{6.91}{13}{25.07}{7.98}{14}
\emline{25.07}{7.98}{15}{22.94}{6.02}{16}
\put(31.00,19.00){\vector(1,0){5.99}}
\put(36.99,19.00){\vector(0,1){11.03}}
\put(36.99,30.03){\vector(-1,0){25.07}}
\put(11.91,30.03){\vector(0,-1){11.03}}
\put(11.91,19.00){\vector(1,0){5.16}}
\put(12.00,7.00){\vector(1,0){5.07}}
\emline{31.00}{7.00}{17}{41.96}{6.91}{18}
\put(42.00,7.00){\vector(1,0){1.03}}
\put(38.00,26.00){\makebox(0,0)[cc]{tuple bit}}
\put(43.00,9.00){\makebox(0,0)[cc]{Copy of tuple}}
\put(10.00,8.00){\makebox(0,0)[cc]{$| 0 >$}}
\end{picture}

}
\caption{Use of controled not gate to copy the data before performing
classical measurement}\label{copying}
\end{figure}

The restriction of only being able to read out one tuple
from the database can be evaded by using controled not gates
as a means of copying the database before measuring it\cite{Deutsch1995}.
By sending the qubits of the tuple through the control input of
a controlled not gate, and qubits prepared in state $ | 0>$ through
the other as shown in {\bf Fig \ref{copying}}, one can create an `oracle' that
acts as a stochastic generator of tuples from the database.
By tailoring the amplitudes of the different tuples in the database one
could tune the probabilities with which they would be read.

Current uses of databases fall into two broad areas, 
transaction processing and management information. 
In the case of the former, the data are bearers of 
important social relations such as relative indebtedness,
and, in consequence, it is of the utmost importance 
that the integrity and detail of the data be preserved.
Were this not the case, there would be a danger that 
alterations of the data would result in changes in
peoples' social status.

In the latter case, the data are used by organisations 
to make decisions about their future courses of action.
Here, the information presented relates not to individual 
people, or individual economic transactions, but to
collections of people and events. One is concerned not 
with what an individual student gained in her A
Levels, but with the mean results in English A 
levels by region of the country, or the average sales of
dishwashers over the last year by month and by model. 
The ultimate source, however, for the summary
information so presented, are the original  transactional 
records demanded by the relations in question.

There is however, an inherent mismatch between the 
transactional sources and the summary uses of 
management information. The sources are voluminous and 
accurate, the uses compact and, although this is
not always appreciated, inherently approximate. 
This approximation arises from two causes. Firstly, the
results presented : totals and averages, are arrived at 
by means of summation, an inherently information
destroying operation. Secondly once one abstracts from 
their individuality, individual commercial
transactions can be seen as stochastic events. 
The ability to directly model this could be an attractive
feature of quantum databases.
 
 \def\root2{\sqrt{2}}
 \begin{figure}\center{
\unitlength=1.00mm
\special{em:linewidth 0.4pt}
\linethickness{0.4pt}
\begin{picture}(69.00,35.00)
\put(24.00,15.00){\framebox(16.00,20.00)[cc]{MIX}}
\put(25.00,31.00){\makebox(0,0)[cc]{A}}
\put(38.00,31.00){\makebox(0,0)[cc]{A'}}
\put(38.00,21.00){\makebox(0,0)[cc]{B'}}
\put(25.00,21.00){\makebox(0,0)[cc]{B}}
\emline{24.00}{30.00}{1}{40.03}{29.97}{2}
\emline{24.00}{20.00}{3}{40.03}{20.08}{4}
\emline{30.00}{30.00}{5}{36.07}{20.08}{6}
\put(18.00,30.00){\vector(1,0){6.01}}
\put(40.00,30.00){\vector(1,0){5.06}}
\put(40.00,20.00){\vector(1,0){5.06}}
\emline{18.00}{20.00}{7}{24.01}{20.08}{8}
\put(18.00,20.00){\vector(1,0){6.01}}
\put(69.00,20.00){\makebox(0,0)[cc]{$B'=\frac{1}{\root2}\mid A> \frac{1}{\root2}\mid B>$}}
\put(62.00,30.00){\makebox(0,0)[cc]{$A'= \mid A >$}}
\end{picture}
 }
 \caption{The MIX gate}
 \label{MIX}
 \end{figure}
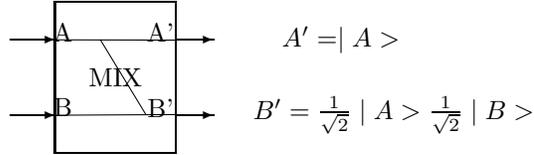
In order to prepare tuples in an appropriate superposition
one needs a primitive operation that will combine the
state of two qubits into one. An operator capble of doing
this would be the MIX gate shown in {\bf Fig. \ref{MIX}}.
This takes two qubits and $A$, and $B$. Bit $A$ passes out unaffected as
$A'$. The second output is an equal mixture of the two input states
$B' = \frac{1}{\root2} (|A>+|B>)$. The MIX gate can
be represented as the matrix:
\begin{equation}
MIX =
\begin{array}{cccc}

 1	&0	&0	&0\\
 \frac{\sqrt{2}}{2} &	\frac{\root2}{2}&	0&	0\\
 0&	0	&\frac{\sqrt{2}}{2}&\frac{\sqrt{2}}{2}\\
 0&	0	&0	&1\\
\end{array}
\end{equation}
It is obviously possible to combine $2^N$ tuples into a superposition
by a MIX network of depth $N$.

The basic operations permited on a relation database are selection, projection and
join \cite{Codd1970}.

Selection forms a new relation $B$ out of all tuples in relation $A$ that meet some
predicate. A particular case of selection uses equality against the {\em primary key}
of the relation, where the primary key is a column which, on its own, uniquely identifies
a tuple.
If primary key selection is performed as a classical operation after quantum measurement
one would need to perform at least $\frac{N}{2}$ operations to have a 50\% chance
of encountering the tuple. 
If, instead of being performed after classical measurement, the operation is
performed in the quantum domain, Grover \cite{Grover1997} has shown that 
primary key selection can be performed in $O(\sqrt{N})$ steps. His technique 
involves repeatedly inverting the phase of the selected word (tuple) and then inverting
the phases of all tuples about their average. The amplitude of the selected tuple
then goes approximately through the sequence 
$$\frac{1}{\sqrt N}, \frac{-1}{\sqrt N},\approx\frac{2}{\sqrt N},\approx\frac{-2}{\sqrt N},\approx\frac{3}{\sqrt N},...
$$
converging to an amplitude of 1 after $\sqrt{N}$ cycles.

For the more general case of a selection which yields a set rather than a singleton
tuple, Grover's algorithm will concentrate the amplitude in $R_p$ the subset of the relation
$R$ that meets predicate $p$ in $O(\sqrt{f})$ steps, where $f$ is the fraction
of tuples meeting the selection criterion.

For primary key selection the quantum search procedure is inferior to the
use of a classical relational database with an index on the primary key, an
operation that costs $O(\log{N})$. For generalised search operations that do
not lend themselves to indexing, it is superior.

Relational database projection can be achieved trivially in the quantum database by simply discarding all
qubits other than those coding for the domains onto which the relation is projected.
Relational projection here translates directly into a projection onto the appropriate
sub-manifold of Hilbert space.

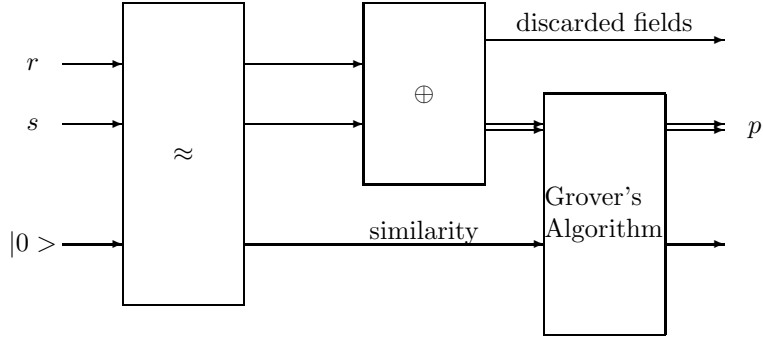
\begin{figure}
\unitlength=0.8mm
\special{em:linewidth 0.4pt}
\linethickness{0.4pt}
\begin{picture}(135.00,60.00)
\put(30.00,10.00){\framebox(20.00,50.00)[cc]{$\approx$}}
\put(20.00,20.00){\vector(1,0){10.00}}
\put(20.00,50.00){\vector(1,0){10.00}}
\put(20.00,40.00){\vector(1,0){10.00}}
\put(50.00,40.00){\vector(1,0){20.00}}
\put(50.00,50.00){\vector(1,0){20.00}}
\put(70.00,30.00){\framebox(20.00,30.00)[cc]{$\oplus$}}
\put(100.00,5.00){\framebox(20.00,40.00)[cc]{\parbox[b]{16mm}{Grover's Algorithm}}}
\put(50.00,20.00){\vector(1,0){50.00}}
\put(90.00,40.00){\vector(1,0){10.00}}
\put(90.00,39.00){\vector(1,0){10.00}}
\put(90.00,54.00){\vector(1,0){40.00}}
\put(120.00,40.00){\vector(1,0){10.00}}
\put(120.00,39.00){\vector(1,0){10.00}}
\put(120.00,20.00){\vector(1,0){10.00}}
\put(15.00,50.00){\makebox(0,0)[cc]{$r$}}
\put(15.00,40.00){\makebox(0,0)[cc]{$s$}}
\put(15.00,20.00){\makebox(0,0)[cc]{$|0>$}}
\put(110.00,57.00){\makebox(0,0)[cc]{discarded fields}}
\put(135.00,39.00){\makebox(0,0)[cc]{$p$}}
\put(80.00,22.00){\makebox(0,0)[cc]{similarity}}
\end{picture}
\caption{The join operation can be performed by composing a similarity operator $\approx$,
         a combining operator $\oplus$ and Grover's algorithm}\label{joinfig}
\end{figure}
\def\rept{{\cal WHILE\ }}
Let $r,s$ be  sets.
Let \begin{equation}\label{join}
p = r(\oplus \join \approx )s
\end{equation}
where \begin{itemize}
        \item$\join$ is the join functional,
        \item $\oplus$ is some dyadic combining operator of type \dyad{$t_r$}{$t_s$}{$t_p$},
        where $t_r$ is the type of the tuples in relation $r$ etc. In
        contemporary relational databases this usually involves some combination of
        permutation and projection on the domains of the two relations, but analytically
        the combining operator can be any function.
        \item $\approx$ is some similarity operator of type 
        \dyad{$t_r$}{$t_s$}{0..1}, two tuples $v,u$ are said to be similar if $(v\approx u)>0$.
        Contemporary relational database systems usually support {\em equijoin} where
        the similarity operator tests fields in the first and second tuple for
        equality, but again, the analytical case is more general and any comparison
        operation is allowed. In conventional databases the result of comparison is
        boolean valued, but that is a special case arising from the fact that 
        a given tuple either is or is not present in the relation. In a quantum
        relation the tuples are present with a complex amplitude, which, on measurement,
        determines the probability of finding the tuple. One can thus see
        two levels of generalisation of the similarity operation:
        \begin{enumerate}
        \item the result of the similarity operator is treated as a real valued 
        classical operator
        such that its quantum realisation is an output in the superposed state
        $(\sqrt{i \approx j})|1>+(\sqrt{1-(i \approx j)})|0>$;
        \item as above but allowing the imaginary component of the amplitudes
        to vary.
        \end{enumerate}
\end{itemize}

then we can say that $p$ contains an element corresponding to every pair of elements
in $x,y$ that are similar.
\begin{equation}\label{sim}
\forall i \in r, \forall j \in s|i\approx j, \exists k\in p   
\end{equation}
each such $k= i\oplus j$ is the result of applying the combining operator to
the pairs of similar elements.

Let us define the conditional similarity $C_{ab}$ of two quantum relations $a,b$
to be :
\begin{equation}\label{csim}
 C_{ab}= \sum_{i\in a}\sum_{j \in b}(i \approx j)\times P_a(i) \times P_b(j) 
\end{equation} Where $P_x(y)$ is the probability of tuple $y$ in relation $x$.
This will be a number in the range 0..1.
We can use it to define the probabilities associated with elements of
a joined set.
Thus in  (\ref{join}) and (\ref{sim}) we have
\begin{equation}\label{joinprob}
P_p (k) = \frac{\sum_{\forall i,j|k=(i\oplus j)} (i \approx j)\times P_a (i) \times  P_b (j)}{C_{rs}}
\end{equation}Note that
\begin{itemize}
        \item
         the probabilities of the elements 
        sum to unity: $\sum_{a\in p}P_p (a) = 1$,
        \item equation (\ref{joinprob}) generalises equation (\ref{sim}).
\end{itemize}

The generalised join operation can be performed as shown in {\bf Fig. \ref{joinfig}}. 
Its complexity is dominated by the Grover's Algorithm network used to boost the
amplitude of the similar joined components, whose complexity will be $O(\frac{1}{\sqrt{C_{rs}}}$.
This contrasts with the complexity of generalised join on a classical computer of
$O(\card r \times \card s)$ where $\card s$ is the cardinality of relation $s$.
If we consider the worst case where the joined relation contains a single tuple,
and 
$C_{rs} = \frac{1}{\card r \times \card s}$, then  the quantum computation takes the
square root of the number of steps of the classical one. Where the conditional similarity
is higher, the complexity advantage of the quantum computation is higher.

In the restricted case of equijoin using a primary key field of relation $s$, the
classical complexity is $O(\card r \times \log{\card s})$. Only when the relation $s$ is
much larger than the relation $r$ and $C_{rs}$ tends to zero,
 does this fall below the complexity of the quantum computation.

The approach given by Grover can be generalised to set an upper complexity limit to
the basic operations of relational databases on a quantum computer. Except in  special cases where
indices can be used on a classical machine, the quantum upper complexity limit is
lower than the classical one.

\end{document}